# 3 Lessons from Hyperinflationary Periods

**Mark Bergen**
Carlson School of Management, University of Minnesota
Minneapolis, MN 55455, USA

**Thomas Bergen**
Carlson School of Management, University of Minnesota
Minneapolis, MN 55455, USA

**Daniel Levy**
Department of Economics, Bar-Ilan University
Ramat-Gan 5290002, Israel
Department of Economics, Emory University
Atlanta, GA 30322, USA
ISET at Tbilisi State University, 0108 Tbilisi, Georgia
ICEA and RCEA

**Rose Semenov**
Carlson School of Management, University of Minnesota
Minneapolis, MN 55455, USA

*Abstract*: Inflation is painful, for firms, customers, employees, and society. But careful study of periods of hyperinflation point to ways that firms can adapt. In particular, companies need to think about how to change prices regularly and cheaply — because constant price changes can ultimately be very, very expensive. And they should consider how to communicate those price changes to customers. Providing clarity and predictability can increase consumer trust and help firms in the long run.

As inflation soars to its highest rates in 40 years, it's critical that businesses have a strategy to respond to fluctuating costs and prices. Pricing during today's inflation is particularly challenging because people are exhausted and emotionally fatigued from dealing with extreme uncertainty created by the ongoing pandemic, the war in Ukraine, and fears of recession.

These stressors affect the fabric of markets and society—amplifying frustration with companies and the overall economy. There are strategies that companies can use to earn consumers' trust during inflation. We have documented these strategies by studying periods of runaway inflation. For example, in Israel during the early 1980s inflation rates rose to over 100% for multiple years, climbing to 430% in 1985 (Snir, Chen, and Levy, 2021). At its extreme, this is known as hyperinflation (inflation of 50% or more per month). The worst hyperinflationary episode in modern history was experienced by Hungary in July 1946, where the inflation rate was 41.9 quadrillion percent, causing prices to double every 15.6 hours (Hanke and Krus, 2013). We've studied how businesses responded to several of these periods, with a particular focus on Israel.

Three major lessons from hyperinflationary periods can help managers, consumers, and societies better cope with and more successfully navigate their current inflationary challenges. Together, they can help companies thrive and lessen the burden that inflation places on consumers as well.

**1. Reduce the company's costs of changing prices ("menu cost")**

Our research shows the costs of changing prices (known is "menu cost") can be far greater than managers recognize, especially during high inflationary environments. We documented the average costs of changing prices per store at four major US grocery chains using in-depth studies of their weekly price change processes (Levy, Bergen, Dutta, and Venable, 1997), analyzing workflow schematics, and undertaking detailed in-store time and motion measurements of each process step across multiple stores at each chain (Levy, Dutta, Bergen, and Venable, 1998). The costs include the labor cost to change shelf prices, the cost of printing and delivering new price tags, the costs of mistakes made during the price change process, and the cost of in-store supervision of the price change process. Applying these costs to the largest US grocery chain, Kroger, would yield an annual cost of **$291.9 million in total** across 2,757 of their stores. Now, imagine a hyperinflationary scenario like Zimbabwe in 2008 where ***prices doubled every day***. These price change processes would be undertaken seven times a week, rather than once a week. This would cause the total annual cost of price adjustment to balloon to **$2.04 billion a year**.

Successfully dealing with high inflation requires managing pricing processes to reduce the costs of price adjustment. This often involves using simplified pricing rules or adopting new pricing technologies. For example, Israeli booksellers went from pricing individually to pricing groups of books by assigning letter codes (A, B, C, …, etc.) and posting a price list. In Brazil, retailers digitized price adjustments with electronic shelf labels, which eliminated the labor costs required to manually update prices across hundreds of products now being re-priced multiple times (rather than once) per week.

The costs of hyperinflation can also be curtailed by grounding prices in more stable currencies. During Israeli runaway inflation, quoting prices in dollars was frequent for durable goods and housing—even at amounts greater than citizens were legally allowed to hold. In Venezuela during the current hyperinflation, sellers are adopting peer-to-peer cryptocurrency payments, allowing street vendors to price with digital coins. By simplifying processes, investing in technology, and presenting price stability, organizations can reduce the costs of changing prices to navigate inflationary pressures.

**2. Reduce customers' "stress of uncertainty"**

During runaway inflation, it is hard to know the value of anything because inflation creates volatile price fluctuations and distorts relative prices. Routine purchases become more complex and demanding, mentally and emotionally. The stress of uncertainty from inflation weighs heavily on customers and is made worse by the uncertainty of the ongoing pandemic, the Ukraine war, and fears of recession. As managers, successfully dealing with high inflation requires focusing on ways to help reduce the customers' stress of uncertainty. This often involves using strategies such as commitment and indexation.

Managers can commit to absorbing some future volatility through offering payment plans like Israeli stores did across many categories during runaway inflation. For example, an Israeli guitar selling for a price of 100 shekels pre-inflation, was advertised at five payments of 100 shekels each during inflation. For customers, this allows them to afford the product immediately even with tighter budgets and finish payments in less valuable future currency. For firms, using payments allowed them to share the risk and burden of hyperinflation with customers at levels supporting economic survival and profitability. Similarly, sellers are using payments to attract customers in Argentina during runaway inflation in categories ranging from furniture to clothing to kitchen utensils to video game consoles.

Indexation allows contract prices to automatically adjust based on the inflation rate. Israeli, German, and Argentinian B2B firms dealt with runaway inflation using indexation to diminish the volatility. During hyperinflation these contracts were indexed to established metrics such as their countries' Consumer Price Index (CPI) or Producer Price Index (PPI). Similar contracting is used by companies buying and selling offerings with volatile commodity or input costs – with the indexation being tied to a particular commodity or input. By focusing on the impact inflation has on your customers, companies can differentiate themselves by helping to solve the everyday challenges of shopping during inflation.

### 3. Reduce societal fears of "shrinkflation, greedflation, and doubtflation"

Inflation is more than just a company or customer problem, it is a societal problem (Moss and Rotemberg, 1998). It affects the fabric of exchange, amplifying frustration with how the economy works. In particular, at times of high inflation customers become suspicious of "shrinkflation" – decreasing product sizes without customers' knowledge – and "greedflation" – companies taking advantage of information and customer exhaustion to increase profitability rather than sharing the burden with customers. Together, these fuel a sense of what we call "doubtflation" – a growing loss in consumer trust in firms and markets as their ability to use prices to make decisions becomes more difficult.

Amid high inflation the onus is on managers to step up to strengthen the credibility of their business decisions and trust with their stakeholders and communities. This involves sharing risks, focusing on relationships, and partnering with employees, supply chain partners, and customers. For example, in Venezuela, a successful retailer shared the burden of hyperinflation with employees by subsidizing prices for essential products, such as access to discounted food at their stores (Cavallo, Cal, and Larangeira (2020). This created substantial savings for employees given the high and rising retail food prices, while paid for at the level of wholesale prices by the company, making it more efficient for the company to deliver. The company also increased trust by communicating weekly with employees, providing information about the company's financial situation, strategy, and the inflation numbers they were using to set prices and wage adjustments.

As another example, one U.S. health care supplier, realizing that passing on pandemic and inflationary costs would be too much for their clinics and hospitals, completely revamped their pricing processes to limit price increases to their providers to allow them to be able to "provide health care." They passed along as much of the cost change as they could when facing this constraint, and raised prices more in places where healthcare provision was not at risk to survive and navigate the extreme cost increase successfully together. Similarly, falafel vendors in Israel chose to not pass on all their current cost increases, taking a short-term loss in order to keep falafels priced as a street food rather than as a luxury. Not all firms can afford this strategy, but it can be a powerful long-term strategy that bolsters trust and cements future business.

Many companies (both B2B and CPG) have increased their partnership efforts with suppliers, customers, communities, and governments to work together to manage inflation. In Israel, to overcome runaway inflation, companies partnered with labor unions and the government to commit to a price and wage freeze for three months as a way to stop the inflationary spiral. This helped eliminate inflation in Israel within a few months.

By aligning the company's pricing strategy around simplifying processes, customers' evolving needs, and society's risks, managers can survive and thrive in even the most difficult inflationary times.